\documentclass[prl,aps,amsmath,superscriptaddress,twocolumn,color]{revtex4-2}
\usepackage{graphicx}
\usepackage{amsmath}
\usepackage{dcolumn}
\usepackage{bm}
\usepackage[T1]{fontenc}

\begin{document}
	
\title{Mixed bubbles in a binary mixture of spin-1 Bose-Einstein condensates}

\author{Uyen Ngoc Le}
\affiliation{Department of Engineering Science, University of Electro-Communications, Tokyo 182-8585, Japan}
\affiliation{Quantum Systems Unit, Okinawa Institute of Science and Technology Graduate University, Onna, Okinawa 904-0495, Japan}

\author{Ha Phuong Uyen Mai}
\affiliation{Department of Engineering Science, University of Electro-Communications, Tokyo 182-8585, Japan}

\author{Hieu Binh Le}
\affiliation{Department of Engineering Science, University of Electro-Communications, Tokyo 182-8585, Japan}

\author{Hiroki Saito}
\affiliation{Department of Engineering Science, University of Electro-Communications, Tokyo 182-8585, Japan}
	
\date{\today}

\begin{abstract}
  In mean-field theory, the ground state of a two-component
  Bose-Einstein condensate (BEC) in a uniform system is either
  uniformly mixed or totally separated.
  Beyond-mean-field effects introduce an additional phase, the
  mixed-bubble state, in which the two components remain mixed within
  one of the separated domains.
  Here we show that a mixed-bubble state can emerge in a binary
  mixture of spin-1 ${}^{7}{\rm{Li}}$ and ${}^{23}{\rm{Na}}$ BECs even
  within the mean-field framework.
  We further demonstrate the existence of a metastable uniformly-mixed
  state, which undergoes a transition to the mixed-bubble state
  triggered by a local perturbation.
\end{abstract}
	
\maketitle
	
Two-component Bose-Einstein condensates (BECs) of ultracold atomic
gases~\cite{Myatt,Ho96} have attracted considerable interest and have
been extensively studied experimentally and theoretically.
A rich variety of phenomena has been explored in these systems,
including phase-separation dynamics~\cite{Hall},
topological excitations~\cite{Leon, Ruo, Seo2015},
vortex lattices~\cite{Kasamatsu, Schweikhard},
coarsening dynamics~\cite{Karl,Hofmann}, and
hydrodynamic instabilities~\cite{Sasaki, Takeuchi, Geng, Huh25}.
Despite this diversity of nonequilibrium and excited-state phenomena,
the ground-state phases of a binary BEC in a uniform system are
remarkably simple within mean-field theory: the system is either
uniformly mixed or totally separated~\cite{Pethick}.
This simplicity originates from the form of the mean-field
interaction-energy density, $\varepsilon(n_1, n_2) = (g_{11} n_1^2 +
g_{22} n_2^2) / 2 + g_{12} n_1 n_2$, where $n_j$ is the density of
component $j$ and $g_{jj'}$ denotes the interaction coefficient
between components $j$ and $j'$.
Since the curvature $c = g_{11} g_{22} - g_{12}^2$ of
$\varepsilon(n_1, n_2)$ is constant, the ground state is uniformly
mixed for $c > 0$ and totally separated for $c < 0$.
No other bulk phase exists within the mean-field
framework~\cite{Pethick}.

By contrast, beyond-mean-field corrections modify the simple quadratic
form of $\varepsilon(n_1, n_2)$ and can generate both positive and
negative curvatures in the energy landscape~\cite{LHY, Petrov}.
As a consequence, a new phase-separated ground state, known as the
mixed-bubble state can emerge~\cite{naidopetrov, sturmer2022}.
In this state, one domain contains both components ($n_1 \neq 0$ and
$n_2 \neq 0$), whereas the other contains only a single component.
However, beyond-mean-field effects become important only near the
miscibility threshold, $g_{11} g_{22} - g_{12}^2 \simeq
0$~\cite{naidopetrov}.
The resulting broad interfaces require large systems to realize
well-developed mixed bubbles.
It is therefore desirable to engineer the energy landscape within
mean-field theory and thereby achieve novel mixing properties,
including mixed-bubble states, without relying on beyond-mean-field
effects.
A possible route was recently proposed using component-dependent
periodic potentials~\cite{Ali2024}.

In this Letter, we show that the energy landscape of a binary mixture
of different atomic species can be engineered by exploiting their
internal spin degrees of freedom.
Unlike scalar BECs, the spin ground states of spinor BECs depend on
$n_1$ and $n_2$, modifying $\varepsilon(n_1, n_2)$ from a simple
quadratic form and thereby generating nontrivial mixing behavior.
Such binary spinor mixtures have been realized
experimentally~\cite{li2015coherent, Eto} and extensively studied
theoretically~\cite{luo2007bose, he2019, he2019_2, xu2009binary,
xu2010spontaneously, shi2010ground, xu2010quantum,
  zhang2010atomic, shi2011three, zhang2011interspecies,
  xu2011quantum, zhang2015fragmentation, xu2012quantum,
  chen2018resonant, jie2021laser, he2020,he2022}.
Their mixing properties, however, remain largely unexplored.
One notable exception is Ref.~\cite{le2024}, which predicted a
metastable uniformly-mixed state in a mixture of spin-1 and spin-2
${}^{87}{\rm Rb}$ condensates.
However, the lifetime of the spin-2 state is limited by
inelastic collisional losses~\cite{Schmaljohann2004, Kuwamoto2004}.
Here we instead focus on a mixture of spin-1 ${}^{7}{\rm{Li}}$ and
${}^{23}{\rm{Na}}$ condensates, whose lowest hyperfine manifolds are
long lived and whose scattering lengths are experimentally
known~\cite{StamperKurn2013, Samuelis, Black, Mil}.
We show that these known interaction parameters are sufficient to
realize mixed-bubble states without any tuning of the scattering
lengths.
This is important because tuning of scattering lengths remains
challenging in spinor BECs.
We obtain the ground-state phase diagram with respect to the density
and external magnetic field and identify a parameter region in which
the mixed-bubble phase emerges.
We also demonstrate the existence of a metastable uniformly-mixed
state and show that a local perturbation can trigger its
transition to the mixed-bubble state.

We consider spin-1 ${}^{7}{\rm{Li}}$ and spin-1 ${}^{23}{\rm{Na}}$
BECs in uniform space at zero temperature.
The system is described within the mean-field approximation by the
macroscopic wave functions $\psi_m^{(j)}(\bm{r})$, where $j=1$ and
2 denote Li and Na, respectively, and $m = 1, 0, -1$ labels the
magnetic sublevels.
We define the density of species $j$ as
$n_j(\bm{r}) = \sum_m |\psi_m^{(j)}(\bm{r})|^2$ and the normalized
spin state as $\zeta_m^{(j)}(\bm{r}) = \psi_m^{(j)}(\bm{r}) /
\sqrt{n_j(\bm{r})}$.
The energy of the system consists of three parts:
$E = E^{(1)} + E^{(2)} + E^{(12)}$.
The energy $E^{(j)}$ of each atomic species $j$ is given
by~\cite{Ohmi, Ho} 
\begin{equation}
  \begin{aligned}[b]
    E^{(j)} = & 
    \int d\bm{r} \sum_m \psi_m^{(j)*}(\bm{r})
    \left[-\frac{\hbar^2}{2M_j} \nabla^2 + \varepsilon_m^{(j)}(B_z)
      \right] \psi_m^{(j)}(\bm{r}) \\
    & + \frac{1}{2} \int d\bm{r} \left[
      g_n^{(jj)} n_j^2(\bm{r}) +
      g_f^{(jj)} \bm{f}^{(j)}(\bm{r}) \cdot \bm{f}^{(j)}(\bm{r})
      \right],
  \end{aligned}
  \label{eq:ej}
\end{equation}
where $M_j$ is the atomic mass, $\varepsilon_m^{(j)}(B_z)$ is the
energy of the sublevel $m$ in the uniform magnetic field $B_z$,
and $\bm{f}^{(j)}(\bm{r}) = \sum_{mm'} \psi_m^{(j)*}(\bm{r})
(\bm{S})_{mm'} \psi_m^{(j)}(\bm{r})$ is the spin density with $\bm{S}$
being the vector of spin-1 matrices.
The interspecies interaction energy $E^{(12)}$ has the
form~\cite{luo2007bose, xu2009binary},
\begin{equation}\label{eq:ejj'}
  \begin{aligned}[b]
    E^{(12)} = &  \int d\bm{r}
    \left[ g_n^{(12)} n_1(\bm{r}) n_2(\bm{r})
      + g_f^{(12)} \bm{f}^{(1)}(\bm{r}) \cdot \bm{f}^{(2)}(\bm{r})
      \right. \\
      & \left. + g_s^{(12)} \left| A_0^{(12)}(\bm{r})  \right|^2 \right],
  \end{aligned}
\end{equation}
where $A_0^{(12)} = (\psi_1^{(1)} \psi_{-1}^{(2)} - \psi_0^{(1)}
\psi_0^{(2)} + \psi_{-1}^{(1)} \psi_1^{(2)}) / \sqrt{3}$ is the
singlet pair density.
The interaction coefficients in Eqs.~(\ref{eq:ej}) and (\ref{eq:ejj'})
are defined as $g_{n, f, s}^{(jj')} = 2\pi\hbar^2 a_{n, f, s}^{(jj')}
/ \mu^{(jj')}$, where $\mu^{(jj')} = M_j M_{j'} / (M_j + M_{j'})$ is
the reduced mass and $a_{n, f, s}^{(jj')}$ is the corresponding
$s$-wave scattering length.
The coupled Gross-Pitaevskii (GP) equations are obtained by the
functional derivative of the total energy as
$i\hbar \partial \psi_m^{(j)} / \partial t = \delta E / \delta
\psi_m^{(j)*}$.

The scattering lengths are known to be $a_n^{(11)} \simeq 12.5 a_B$
and $a_f^{(11)} \simeq -5.7 a_B$ for
${}^{7}{\rm{Li}}$~\cite{StamperKurn2013},
$a_n^{(22)} \simeq 51.1 a_B$ and $a_f^{(22)} \simeq 0.82 a_B$ for
${}^{23}{\rm{Na}}$~\cite{Samuelis, Black}, and
$a_n^{(12)} \simeq 19.65 a_B$, $a_f^{(12)} \simeq 0.35 a_B$, and
$a_s^{(12)} \simeq 0$ between ${}^{7}{\rm{Li}}$ and
${}^{23}{\rm{Na}}$~\cite{Mil}, where $a_B$ is the Bohr radius.
Since $a_f^{(11)}$ is negative and its magnitude is comparable to
$a_n^{(11)}$, ${}^{7}{\rm{Li}}$ is strongly
ferromagnetic~\cite{Huh}.
Consequently, the ground state for zero magnetic field is trivial:
a totally-separated phase consisting of ferromagnetic
${}^{7}{\rm{Li}}$ and polar ${}^{23}{\rm{Na}}$ domains [see Sec.~I of
  Supplemental Material (SM)~\cite{SM} for the miscibility
  of these spin states].
To induce nontrivial mixing properties, we therefore apply a uniform
magnetic field $B_z$ along the $z$ direction.

In a single spinor condensate, the linear Zeeman term can be
eliminated by a unitary transformation, leaving the quadratic Zeeman
effect as the relevant magnetic-field contribution.
In the present heteronuclear mixture, however, such a simplification
is not possible because the magnetic moments of ${}^{7}{\rm{Li}}$ and
${}^{23}{\rm{Na}}$ differ slightly.
We therefore incorporate the magnetic-field dependence using the
Breit-Rabi formula~\cite{BreitRabi1931},
\begin{equation} \label{BR}
  \varepsilon_m^{(j)}(B_z) = -\frac{\Delta E_\text{hf}^{(j)}}{8}
  - \frac{2}{3} \mu_n^{(j)} m B_z
  - \frac{\Delta E_\text{hf}^{(j)}}{2} \sqrt{1 + m x + x^2},
\end{equation}
where $\Delta E^\text{(1)}_\text{hf}/h \simeq 0.804$ GHz and
$\Delta E^\text{(2)}_\text{hf}/h \simeq 1.772$ GHz are the hyperfine
splittings, $\mu_n^{(1)} = 3.256 \mu_N $ and $\mu_n^{(2)} = 2.218 \mu_N$
are the nuclear magnetic moments, and
$x = 2 (\mu_B + \mu_n^{(j)} / 3) B_z / \Delta E_\text{hf}^{(j)}$ with
$\mu_B$ and $\mu_N$ being the Bohr magneton and nuclear magneton,
respectively.
For the magnetic-field range $\sim$ G considered here, no magnetic
Feshbach resonance exists in the ${}^{7}{\rm{Li}}$-${}^{23}{\rm{Na}}$
mixture~\cite{Hulet, Abeelen, Gacesa}.

We investigate the ground state in the thermodynamic limit, where the
interfacial energy is negligible compared with the bulk energy and
each domain may be regarded as uniform.
The system is then characterized by the spin state $\zeta_m^{(j)}$ in
each domain, the domain volume fraction, and the distribution of atoms
between the domains (see Sec.~II of SM~\cite{SM} for details).
For given average densities $\bar{n}_j \equiv N_j / V$ and magnetic
field $B_z$, we numerically minimize the total energy with respect to
these parameters, where $N_j$ is the number of atoms for species $j$
and $V$ is the volume of the system.
Throughout this work, we restrict ourselves to the subspace
\begin{equation} \label{Fz0}
  F_z^{\rm total} = \int [f_z^{(1)}(\bm{r}) + f_z^{(2)}(\bm{r})]
  d\bm{r} = 0,
\end{equation}
which corresponds to experimental situations in which the system is
prepared with zero total magnetization and subsequently relaxes toward
its ground state while conserving $F_z^{\rm total}$~\cite{Stenger}.

\begin{figure}[t]
\includegraphics[scale=0.4]{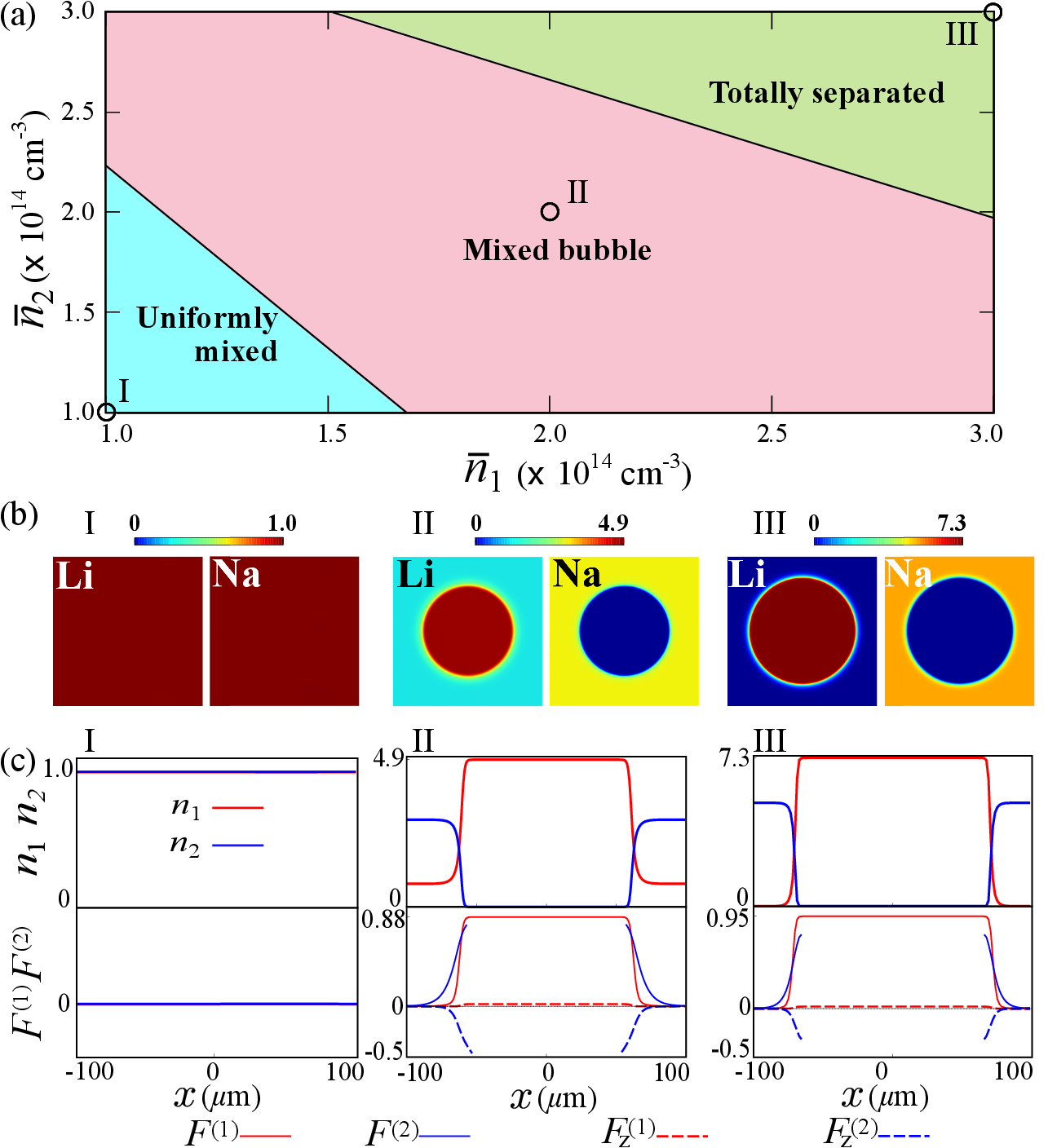} 
\caption{
  Ground-state properties of a binary mixture of spin-1
  ${}^{7}{\rm{Li}}$ and ${}^{23}{\rm{Na}}$ BECs for $B_z = 2$ G.
  (a) Ground-state phase diagram with respect to the average densities
  $\bar{n}_1 = N_1 / V$ and $\bar{n}_2 = N_2 / V$.
  The circles labeled I-III indicate the parameters used in the lower
  panels.
  (b) Two-dimensional density distributions obtained by solving the GP
  equation.
  (c) One-dimensional density profiles $n_j(x, y=0)$ and magnetization
  profiles $F^{(j)}(x, y=0)$, where the origin is located at the
  center.
  States I, II, and III correspond to the uniformly-mixed,
  mixed-bubble, and totally-separated phases, respectively.
}
\label{fig:phaseDia}
\end{figure}

Figure~\ref{fig:phaseDia}(a) presents the ground-state phase diagram
for $B_z = 2$ G.
Three distinct phases are found: the uniformly-mixed, mixed-bubble,
and totally-separated phases.
This is in stark contrast to the zero-field case, where only the
totally-separated phase exists.
For large densities $\bar{n}_1$ and $\bar{n}_2$, the interaction
energy dominates over the magnetic-field energy, and the two species
remain totally separated, similarly to the $B_z = 0$ case.
At low densities, by contrast, the magnetic-field contribution becomes
relatively important.
The quadratic Zeeman effect then favors the polar state for both
species, leading to a uniformly-mixed phase (see Sec.~I of
SM~\cite{SM}).
Between these two limits, a mixed-bubble phase emerges.
As shown below, the spin ground states depend nonlinearly on the
mixing ratio, deforming the energy landscape from a simple quadratic
form to one containing both convex and concave regions.
This modification stabilizes the mixed-bubble phase.

Representative density profiles obtained by imaginary-time evolution
of the GP equation are also shown in Fig.~\ref{fig:phaseDia}.
The dimensionality is unimportant for the mixing properties, and we
present two-dimensional results for simplicity~\cite{2D}.
In both the mixed-bubble and totally-separated phases, the system
separates into an inner and an outer domain.
In the mixed-bubble phase, ${}^{7}{\rm{Li}}$ occupies both domains,
and one of the domains contains both atomic species, as shown in
Fig.~\ref{fig:phaseDia}(b).
By contrast, in the totally-separated phase each domain contains only
a single atomic species.
The spin state in each domain can be characterized by the
magnetization $\bm{F}^{(j)} = \sum_{mm'} \zeta_m^{(j)*} (\bm{S})_{mm'}
\zeta_{m'}^{(j)}$.
In the mixed-bubble phase, the domain containing only
${}^{7}{\rm{Li}}$ exhibits a large magnetization $F^{(j)} =
|\bm{F}^{(j)}|$ owing to the strong ferromagnetic interaction of
${}^{7}{\rm{Li}}$, whereas the magnetization of ${}^{7}{\rm{Li}}$ is
strongly suppressed in the mixed domain by the antiferromagnetic
interspecies spin interaction ($a_f^{(12)} > 0$).

\begin{figure}[t]
\includegraphics[scale=0.6]{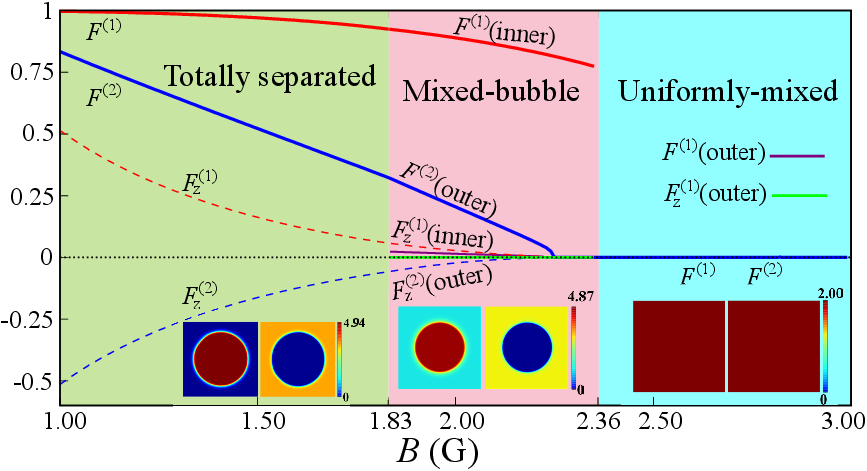} 
\caption{
  Ground-state phases as a function of the external magnetic field
  $B_z$ for fixed average densities $\bar{n}_1 = \bar{n}_2 = 2 \times
  10^{14}$ ${\rm cm}^{-3}$.
  Shown are the magnetizations $F^{(j)} = |\bm{F}^{(j)}|$ and
  $F_z^{(j)}$ of the ground-state domains.
  In the mixed-bubble phase, ${}^{7}{\rm{Li}}$ occupies both the mixed
  and pure-${}^{7}{\rm{Li}}$ domains, resulting in two branches for
  $F^{(1)}$ and $F_z^{(1)}$.
  The upper branches terminate at $B_z \simeq 2.36$ G, where the
  pure-${}^{7}{\rm{Li}}$ domain vanishes.
  The insets show representative GP density profiles for $B_z = 1.5$,
  2.0, and 2.5 G.
}
\label{changeB}
\end{figure}

Figure~\ref{changeB} shows the ground-state phases as a function of
the magnetic field $B_z$ for fixed average densities $\bar{n}_1 =
\bar{n}_2 = 2 \times 10^{14}$ ${\rm cm}^{-3}$.
At low magnetic fields, the interaction energy dominates and the
system is totally separated, consistent with
Fig.~\ref{fig:phaseDia}(a).
As $B_z$ increases, the system undergoes a transition to the
mixed-bubble phase and eventually to the uniformly-mixed phase.
In the totally-separated phase, both ${}^{7}{\rm{Li}}$ and
${}^{23}{\rm{Na}}$ are in broken-axisymmetry
states~\cite{Murata, Note}.
This feature persists throughout the mixed-bubble region above $B_z
\simeq 1.83$ G.
In this regime, the spin states in all domains are broken-axisymmetry
states, although the magnetization of ${}^{7}{\rm{Li}}$ is strongly
suppressed in the mixed domain.
Above $B \simeq 2.36$ G, the uniformly-mixed phase consists of polar
states for both atomic species.


To elucidate the mechanism underlying the mixed-bubble phase, we
temporarily consider the Gibbs ensemble at fixed pressure $P$.
Although the system of interest is characterized by fixed volume, the
fixed-pressure ensemble yields the same qualitative conclusions while
providing a simpler physical interpretation.
To simplify the problem further, we neglect the effect of the nuclear
magnetic moments in Eq.~(\ref{BR}) and only take the quadratic Zeeman
energies into account as $\varepsilon_m^{(j)}(B_z) = q_j m^2$, where
$q_j = \mu_B^2 B_z^2 / (4 \Delta E^\text{(j)}_\text{hf})$ is the
quadratic Zeeman coefficient.
For simplicity, we approximate the spin state of ${}^{23}{\rm{Na}}$ by
the polar state $\bm{\zeta}^{(2)} = (0,1,0)$.
Under these assumptions, the spin ground state for ${}^{7}{\rm{Li}}$
with $F_z^{(1)} = 0$ is given by $\bm{\zeta}^{(1)} = [\sin(\theta / 2)
  / \sqrt{2}, \cos(\theta / 2), \sin(\theta / 2) / \sqrt{2}]$, where
$\theta = \cos^{-1} (q_1 / q_{\rm cr})$ for $q_1 < q_{\rm cr}$ and
$\theta = 0$ for $q_1 > q_{\rm cr}$ with $q_{\rm cr} = 2 |g_f^{(11)}|
n_1$~\cite{Stenger}.
The Gibbs free energy per particle, $g \equiv (E + PV) / (N_1 + N_2)$,
of the uniformly-mixed state for these spin states is given by (see
Sec.~III of SM~\cite{SM} for derivation)
\begin{eqnarray} \label{G1}
  g & = & 2 \left( P + \frac{q_1^2}{8 g_f^{(11)}} \right)^{1/2}
  \Biggl[ \frac{g_n^{(11)} + g_f^{(11)}}{2} R^2
  \nonumber \\
  & & + \frac{g_n^{(22)}}{2} (1-R)^2 + g_n^{(12)} R(1-R) \Biggr]^{1/2}
    + \frac{q_1}{2} R
\end{eqnarray}
for $q < q_{\rm cr}$ and
\begin{equation}
\label{G2}
g = 2 P^{1/2} \left[ \frac{g_n^{(11)}}{2} R^2
  + \frac{g_n^{(22)}}{2} (1-R)^2 + g_n^{(12)} R (1 - R) \right]^{1/2}
\end{equation}
for $q \geq q_{\rm cr}$, where $R = N_1 / (N_1 + N_2)$.

The shape of $g(R)$ determines both the energetic stability of the
uniformly-mixed state against phase separation and the resultant
equilibrium phase~\cite{Porter}.
Suppose that a uniformly-mixed state with composition $R$ separates
into two domains with compositions $R_a < R$ and $R_b > R$.
The Gibbs free energy changes from $g(R)$ to
$g_{\rm sep} = [(R_b - R) g(R_a) + (R - R_a) g(R_b)] / (R_b - R_a)$,
which is geometrically represented by the intersection between the
vertical line at $R$ and the line segment connecting $[R_a, g(R_a)]$
and $[R_b, g(R_b)]$ (see Sec.~IV of SM~\cite{SM} for details).
Therefore, $g''(R) < 0$ indicates instability against phase
separation, whereas $g''(R) > 0$ implies local stability of the
uniformly-mixed state.

\begin{figure}[t]
  \includegraphics[width=8.8cm]{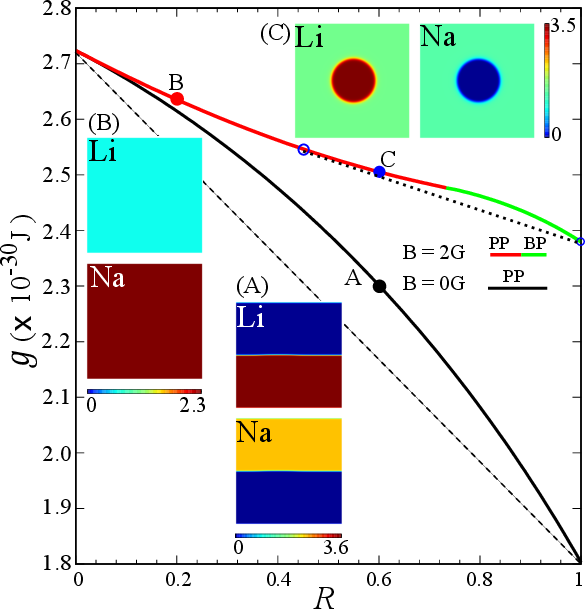} 
  \caption{
    Gibbs free energy per particle $g$ in Eqs.~(\ref{G1}) and
    (\ref{G2}) as a function of the composition $R = N_1 / (N_1 +
    N_2)$ for $B_z = 0$ and 2 G at fixed pressure $P = 3.75 \times
    10^{-10}$ ${\rm N} / {\rm m}^2$.
    The abbreviations PP and BP denote states in which
    ${}^{7}{\rm{Li}}$ is in the polar and broken-axisymmetry states,
    while ${}^{23}{\rm{Na}}$ is assumed to remain in the polar state.
    Insets show the ground states obtained by solving the GP equation
    for (A) $R = 0.6$ and $B = 0$, (B) $R = 0.2$ and $B = 2$ G,
    and (C) $R = 0.6$ and $B = 2$ G.
    In the GP simulation, the total number of atoms is adjusted to
    give the same $P$ as that used in the main panel.
    The dashed line indicates the lowest-energy totally-separated
    state, and the dotted line the common tangent corresponding to the
    mixed-bubble state.
    The open circles mark the compositions of the coexisting domains.
  }
  \label{fig:nali}
\end{figure}

Figure~\ref{fig:nali} shows $g(R)$ in Eqs.~(\ref{G1}) and (\ref{G2})
for $B_z = 0$ and 2 G.
For $B_z = 0$, $g(R)$ is concave over the entire range of $R$,
indicating that the uniformly-mixed state is always unstable against
phase separation (see Fig.~S1(b) in SM~\cite{SM}).
The lowest-energy state is obtained from the common tangent connecting
$R_a = 0$ and $R_b = 1$ (dashed line in Fig.~\ref{fig:nali}),
corresponding to the totally-separated phase.
By contrast, for $B_z = 2$ G, $g(R)$ develops a convex-concave
structure.
The common-tangent construction then predicts a mixed-bubble state
consisting of domains with $R_a \simeq 0.3$ and $R_b = 1$, indicated
by the open circles in Fig.~\ref{fig:nali}.
The GP simulations shown in the insets are in good agreement with
these predictions.
The origin of the mixed-bubble phase is thus clear.
As the composition $R$ changes, the spin ground state of
${}^{7}{\rm{Li}}$ undergoes a polar-to-broken-axisymmetry transition
(PP-BP transition in Fig.~\ref{fig:nali}).
The resulting change in the spin interaction energy deforms $g(R)$
from a purely concave curve into a convex-concave one, thereby
stabilizing the mixed-bubble phase.
Notably, $g(R)$ remains locally convex around $R = 0.6$ (point C in
Fig.~\ref{fig:nali}), suggesting that the uniformly-mixed state can
survive as a metastable state even though the true ground state is the
mixed-bubble state.

\begin{figure}[t]
\includegraphics[width=8.8cm]{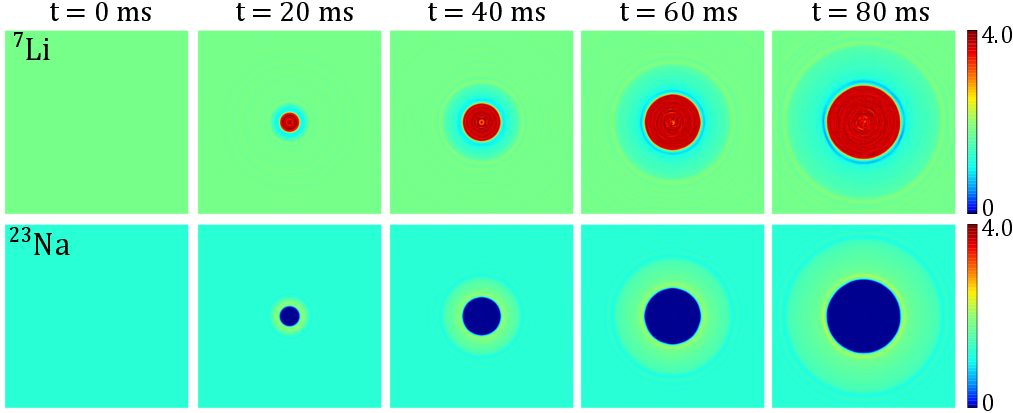} 
\caption{
  Real-time evolution obtained by solving the GP equation for $B_z =
  2$ G.
  The initial state is a uniformly-mixed state in which both
  ${}^{7}{\rm{Li}}$ and ${}^{23}{\rm{Na}}$ are in the polar state,
  with $n_1 = 1.8 \times 10^{14}$ ${\rm cm}^{-3}$
  and $n_2 = 1.2 \times 10^{14}$ ${\rm cm}^{-3}$
  A local perturbation [Eq.~(\ref{dist})] is applied with $B_0 = 1.6$
  mG and $a = 4$ $\mu{\rm m}$.
  The perturbation nucleates a pure-${}^{7}{\rm{Li}}$ domain near the
  center, which triggers a transition to the mixed-bubble state
  throughout the system.
}  
\label{fig:PPdynamic}
\end{figure}

To confirm this metastability, we numerically solve the real-time
GP equation starting from a uniformly-mixed state in which both
${}^{7}{\rm{Li}}$ and ${}^{23}{\rm{Na}}$ are in the polar state.
To induce phase separation, we add a local perturbation,
\begin{equation} \label{dist}
  \frac{\mu_B}{2} B_0 e^{-(x^2+ y^2) / a^2} \sum_{m'}
  (f_x^{(1)})_{mm'} \psi_{m'}^{(1)}(\bm{r})
\end{equation}
to the GP equation for ${}^{7}{\rm{Li}}$.
Such a synthetic local magnetic field can be realized using laser
beams~\cite{Grimm2000}.
Figure~\ref{fig:PPdynamic} shows the resulting dynamics.
The local perturbation nucleates a pure ${}^{7}{\rm{Li}}$ domain at
the center, which subsequently grows and triggers a transition to the
mixed-bubble phase throughout the system.
We have verified that sufficiently small perturbations do not trigger
the transition.
The initial uniformly-mixed state is therefore metastable rather than
dynamically unstable.

In conclusion, we have shown that a mixed-bubble phase can emerge in a 
binary mixture of spin-1 ${}^{7}{\rm{Li}}$ and ${}^{23}{\rm{Na}}$
Bose-Einstein condensates within mean-field theory.
The mechanism originates from the composition dependence of the spin
ground states, which deforms the energy landscape from a simple
quadratic form into one containing both convex and concave regions.
We obtained the ground-state phase diagram, identified the parameter
regime supporting the mixed-bubble phase, and demonstrated the
existence of a metastable uniformly-mixed state whose transition to
the mixed-bubble state can be triggered by a local perturbation.

This work was supported by JSPS KAKENHI Grant Numbers JP23K03276 and
JP26K00638.

%
	

\begin{thebibliography}{99}
\bibitem{2D}
  A two-dimensional model corresponds to the experimental situation
  in which the system is confined in a quasi-two-dimensional box
  potential whose size is much larger than domain sizes.
  The three-dimensional interaction coefficients can still be used,
  when the width of the box is much larger than the healing length.
\bibitem{Note}
  Although an isolated spin-1 ${}^{23}{\rm{Na}}$ condensate has a
  polar ground state, the finite value of $F_z^{(2)}$ in the
  totally-separated phase originates from the constraint in
  Eq.~(\ref{Fz0}) and the species dependence of the linear Zeeman
  energy in Eq.~(\ref{BR}).
\end{thebibliography}
\end{document}